\documentclass{biophys}
\usepackage{helvet,times}
\usepackage{bm,textcomp}

\jno{kxl014} 
\gridframe{N}
\cropmark{N}

\usepackage{amsmath}

\usepackage[round,numbers,sort&compress]{natbib}
\begin{document}

%
%
%
%
%

\title{Stabilizing membrane domains antagonizes \textcolor{black}{n-alcohol} anesthesia}

\author{B.B. Machta, E. Gray, M. Nouri, N.L.C. McCarthy,\\ E.M. Gray, A.L. Miller, N.J. Brooks and S.L. Veatch }


\maketitle 
\clearpage

{\large
\noindent \textbf{ABSTRACT}  %

{Diverse molecules induce general anesthesia with potency strongly correlated both with their hydrophobicity and their effects on certain ion channels. We recently observed that several \textcolor{black}{n-alcohol} anesthetics inhibit heterogeneity in plasma membrane derived vesicles by lowering the critical temperature ($T_c$) for phase separation. Here we exploit conditions that stabilize membrane heterogeneity to \textcolor{black}{further} test the correlation between the anesthetic potency of n-alcohols and effects on $T_c$. First we show that hexadecanol acts oppositely to \textcolor{black}{n-alcohol} anesthetics on membrane mixing and antagonizes ethanol induced anesthesia in a tadpole behavioral assay. Second, we show that two previously described `intoxication reversers' raise $T_c$ and counter ethanol's effects in vesicles, mimicking the findings of previous electrophysiological \textcolor{black}{and behavioral} measurements.  Third, we find that hydrostatic pressure, long known to reverse anesthesia, also raises $T_c$ in vesicles with a magnitude that counters the effect of \textcolor{black}{butanol} at relevant concentrations and pressures.  Taken together, these results demonstrate that $\Delta T_c$ predicts anesthetic potency \textcolor{black}{for n-alcohols} better than hydrophobicity in a range of contexts, supporting a mechanistic role for membrane heterogeneity in general anesthesia.}
\linebreak

\maketitle 

\noindent \textbf{INTRODUCTION}

The potencies of many general anesthetics are roughly proportional to their oil:water partition coefficient over more than five orders of magnitude in overall concentration~\citep{Meyer99}.  This Meyer-Overton correlation suggests membrane involvement, and anesthetics have been shown to decrease lipid chain ordering, lower the main chain melting temperature, and increase membrane spontaneous curvature, fluidity, and conductance~\citep{Gruner91,Wodzinska09,Ingolfsson11}.  However, these effects are small~\citep{Herold14} and often cannot account for those molecules which deviate from Meyer-Overton~\citep{Franks86}. 
 Most recent attention focuses on the ion channels known to be most sensitive to these compounds~\citep{Franks94}, where extensive structural work~\citep{Mihic97,Borghese06,Nury11} suggests that anesthetic effects are mediated by specific residues in hydrophobic, membrane spanning regions.   \textcolor{black}{High potency general anesthetics such as etomidate and barbiturates are more effective at potentiating channels and producing anesthesia than predicted from their hydrophobicity alone, and evidence is accumulating that these compounds bind directly and specifically to channels at the interface between subunits~\citep{Forman16}.  Many researchers also favor a direct binding mechanism of anesthetic action for lower potency anesthetics, including the n-alcohol anesthetics investigated here~\cite{Mihic97}.  However, channels are proposed to contain more numerous binding sites for these compounds~\citep{Xu00,Brannigan09,Forman16}, each with low affinity, leaving open the possibility that these anesthetics interact with channels more as a solvent than as a ligand.}

Our understanding of the structure and function of the animal plasma membrane has grown dramatically since most membrane theories of anesthesia were put forward. It is now appreciated that animal plasma membranes have a thermodynamic tendency to separate into coexisting liquid domains, sometimes referred to as `lipid rafts' or `lipid shells' \cite{Simons97,Anderson02} and that this heterogeneity localizes and regulates ion channels, sometimes in a subtype specific manner \cite{Allen07}.
Much of this regulation likely arises from the membrane's unusual thermodynamic properties.   Cholesterol containing membranes of purified lipids can support two distinct liquid phases~\citep{Veatch05}, and giant plasma membrane vesicles (GPMVs) isolated from mammalian cell lines display analogous 
phase coexistence at low temperature~\citep{Baumgart07}.  Remarkably, GPMVs are near the critical point of this transition~\citep{Veatch08}, a non-generic region of phase space distinguished by large correlation times and \textcolor{black}{large but finite} domain sizes that requires fine tuning both composition and temperature in synthetic systems~\citep{HonerkampSmith08}.

We recently found~\citep{Gray13} that incubating several general anesthetics with isolated GPMVs lowered their critical temperatures ($T_c$) in a way that scales well with their anesthetic dose as previously measured in tadpole loss of righting reflex (LRR) assays~\citep{Pringle81}. 
While our assay measures a change in $T_c$ somewhat below growth temperature, we predict that at higher temperatures these treatments would destabilize sub-micron liquid domains both in vesicles and the intact plasma membranes from which they were derived.
While suggestive, we wanted to rule out the possibility that our observed correlation is derivative of a more fundamental correlation of both $\Delta T_c$ and anesthetic potency with hydrophobicity.
 As a first step towards this, we demonstrated that two hydrophobic but non-anesthetic analogs of general anesthetics did not affect $T_c$ at concentrations where Meyer-Overton predicts they would~\citep{Gray13}. 
Here we explore a more direct challenge to the connection between changes in $T_c$ and anesthetic potency by investigating the anesthetic effects of hydrophobic compounds and conditions that \textit{raise} transition temperatures in GPMVs. 
\linebreak
\linebreak
\noindent \textbf{MATERIALS AND METHODS}

\noindent \textit{Giant plasma membrane vesicle (GPMV) measurements:}
RBL-2H3 cells \cite{Barsumian81}  were maintained in MEM media with 20\% FBS and 0.1\% Gentamycin at 37$^\circ$C in 5\% CO$_2$. XTC-2 cells \cite{Pudney73} were maintained in L-15 Media diluted 1:1.5 with water for amphibian cells with 10\% FBS, sodium bicarbonate (2.47 g/liter), pen strep (100 units/ml) at room temperature in 5\% CO$_2$. Freshly seeded cells were incubated in complete media for at least 18h at the growth temperature indicated prior to GPMV isolation.   All culture reagents were purchased from Fisher Scientific (Hampton, New Hampshire). Other reagents were purchased from Sigma Aldrich (St. Louis, MO) at the highest available purity unless otherwise indicated.

GPMVs from RBL cells were prepared through incubation with low concentrations of dithiothreitol (DTT, $2$mM) and formaldehyde ($25$mM) in the presence of calcium ($2$mM) for 1h as described previously \cite{Gray13}. \textcolor{black}{GPMVs isolated from RBL cells contain 20-40 mol$\%$ cholesterol, sphingolipids, phospholipids, and gangliosides typically associated with the plasma membrane, along with many transmembrane and peripheral plasma membrane proteins~\citep{Fridriksson99, Baumgart07,Levental16}.  }
 For XTC-2 derived GPMVs, the vesiculation buffer was diluted 1:5 while maintaining calcium, formaldehyde, and DTT concentrations, and cells were incubated for at least 2h at room temperature. Prior to GPMV formation, cells were labeled with DiI-C12 (Life Technologies, Carlsbad, CA; 2$\mu$g/ml in 1\% methanol) for 10min at room temperature.  GPMVs probed at atmospheric pressure were imaged on an inverted microscope (IX81; Olympus, Center Valley, PA) with a 40x air objective (0.95 NA), epi-illumination using an Hg lamp and Cy3 filter set (Chroma Technology, Bellows Falls, VT).  Temperature was controlled using a home built Peltier stage described previously \cite{Gray13} coupled to a PID controller (Oven Industries, Mechanicsburg, PA), and images were recorded using a sCMOS camera (Neo; Andor, South Windsor, CT).

GPMV suspensions with hexadecanol were prepared either using super-saturated solutions or equilibrated solutions.  To make super-saturated solutions, hexadecanol was suspended in either DMSO or ethanol using volumes corresponding to the final concentration indicated in figures, then mixed directly with the GPMV suspension in aqueous buffer while mixing.  The maximum DMSO concentration used was 0.5\% v/v, and previous work demonstrates that $T_c$ is not affected by DMSO treatment alone ~\cite{Gray13}.  Equilibrated solutions were prepared by first adding a concentrated hexadecanol stock directly to the GPMV suspension such that it precipitated out of solution.  Then, the desired volume of ethanol was added and the solution mixed to facilitate the resuspension of hexadecanol.  \textcolor{black}{In all cases where ethanol and hexadecanol were added in combination, they were maintained at a 120mM ethanol to 30$\mu$M hexadecanol ratio, as indicated in figures.  Equilibrated solutions could be made for the following: 60mM:1.5$\mu$M, 120mM:3$\mu$M, 180mM:4.5$\mu$M, and 240mM:6$\mu$M (ethanol:hexadecanol), although for higher concentrations some small fraction of hexadecanol lacked solubility.  Only super-saturated solutions could be made for 600mM ethanol:15$\mu$M hexadecanol. }
\linebreak

\begingroup
    \includegraphics[width=3.25in]{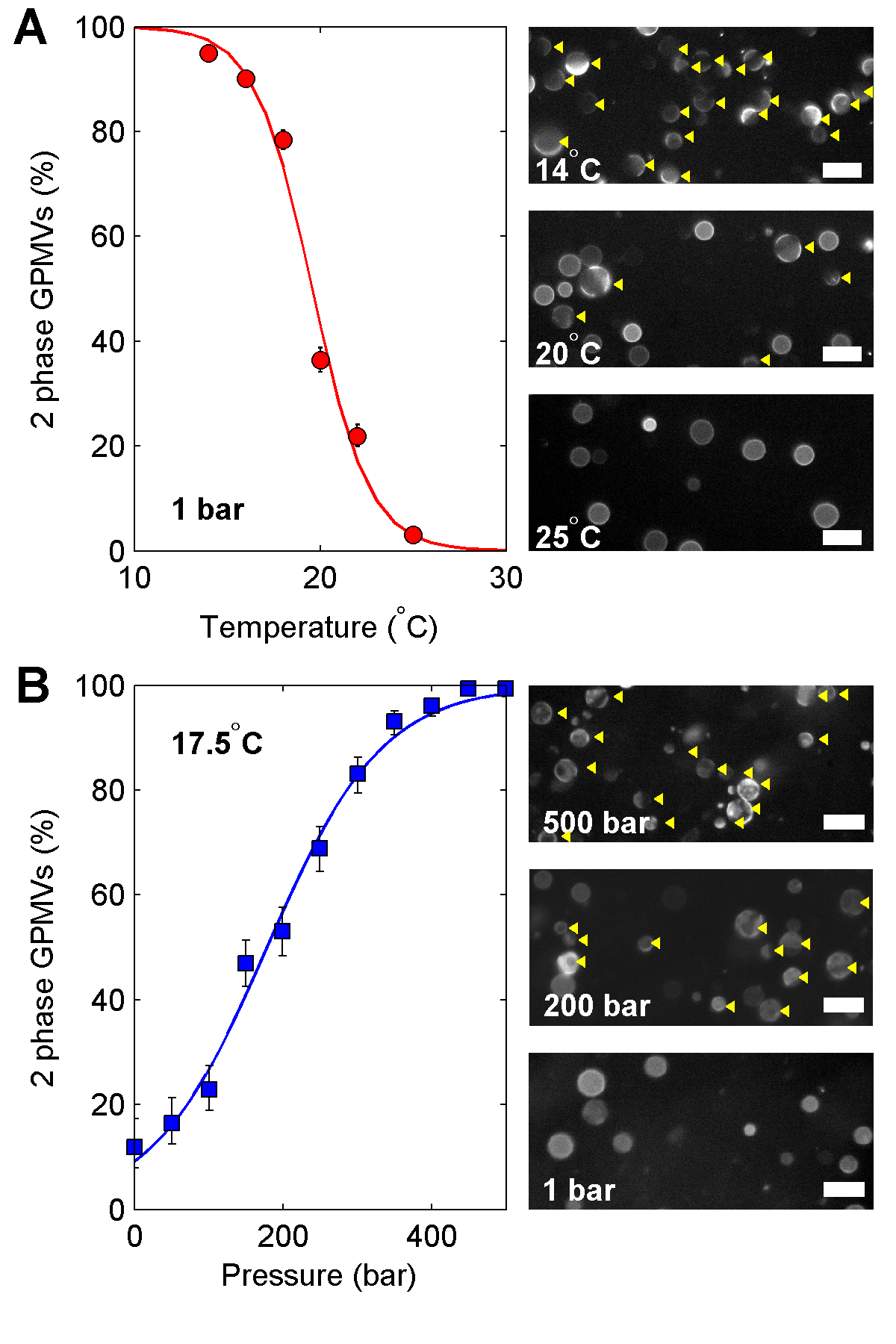}
		
\normalsize{
\textcolor{black}{
\noindent {\bf Fig. 1.} Determination of the average critical temperature or pressure of DiIC$_{12}$ labeled GPMVs through fluorescence imaging. (A) Fields containing multiple GPMVs were imaged over a range of temperatures and at fixed pressure, with representative subsets of images shown on the left.  At high temperature, most GPMVs appear uniform, while an increasing fraction of vesicles appear phase separated as temperature is lowered, with phase separated vesicles indicated by yellow arrows.  We manually tabulate the fraction of GPMVs that contain two coexisting liquid phases as a function of temperature from these images, constructing the plot on the right.  These points are fit to the sigmoid function described in Methods to determine the extrapolated temperature where 50$\%$ of vesicles contain coexisting liquid phases.  (B) Fields containing multiple GPMVs were imaged over a range of pressures at fixed temperature, and representative subsets of images are shown on the left.  At low pressure, most GPMVs appear uniform, while an increasing fraction of vesicles appear phase separated as pressure is increased. As with the fixed pressure data in A, these points are fit to the sigmoid function described in Methods to determine the extrapolated pressure where 50$\%$ of vesicles contain coexisting liquid phases.  
}
\linebreak
\linebreak
}
\endgroup

Measurements conducted at elevated hydrostatic pressures were made using a custom built microscopy compatible pressure cell mounted on a Nikon Eclipse TE2000-E inverted microscope as described previously \cite{McCarthy15,Purushothaman15} with a 20x extra long working distance air objective (0.4 NA) and G2-A filter set (Nikon Instruments, Richmond, UK) The pressure cell temperature was controlled via a circulating water bath. Images were acquired using a sCMOS based camera (Zyla; Andor, Belfast, UK) and recorded using custom built software with temperature and pressure logging.

GPMV transition temperatures at constant pressure were measured as described previously \cite{Gray13} \textcolor{black}{and as illustrated in Fig 1A.}  Briefly, images were acquired of fields of GPMVs over a range of temperatures such that at least 100 vesicles were detected at each temperature. After imaging, individual vesicles were identified as having a single liquid phase or two coexisting liquid phases.  This information was compiled into a plot showing the percentage of vesicles with two liquid phases as a function of temperature, which was fit to a sigmoid function to extrapolate the temperature where 50\% of vesicles contained two coexisting liquid phases,  
$\textrm{\% Phase Separated} = 100 \times\left(1- \frac{1}{1+e^{-(T-T_c)/B}}\right)$,
where $B$ is a parameter describing the width of the transition. \textcolor{black}{The width of the transition within a population of vesicles ($\sim10^\circ$C) is much broader than the width of the transition for a single GPMV ($<2^\circ$C)~\citep{Veatch08}, most likely due to heterogeniety in composition between GPMVs~\cite{Gray15}. }

 We have previously demonstrated that these GPMVs pass through a critical temperature at the transition, even in the presence of anesthetics, therefore we refer to this temperature as the average critical temperature ($T_c$) of the sample.  Errors in single measurements of $T_c$ ($\sigma_{T_c}$) are 68\% confidence interval estimates of this parameter determined directly from the fit.  \textcolor{black}{For the example shown in Fig 1A, the critical temperature is $T_c$=19.5$\pm$0.6$^\circ$C. We generally observe that average critical temperatures measured in this way vary between 12$^\circ$C $<T_c<$ 27$^\circ$C for untreated RBL-derived GPMVs prepared as described above, depending on the growth conditions of the cells from which the GPMVs were derived~\citep{Gray15}.  XTC-2 cells produced GPMVs with average critical temperatures of 14.8$\pm$0.4$^\circ$ and 20.7$\pm$0.4$^\circ$C. } Error bounds for a transition temperature shift ($\Delta T_c$) are given by $\sqrt{\sigma^2_M+\sigma^2_C}$ where $\sigma_C$ is the error in measuring $T_c$ of the untreated control and $\sigma_M$ is the error in measuring $T_c$ of the treated sample. The average critical pressure in the sample ($P_c$) at constant temperature was determined by a similar procedure, \textcolor{black}{as illustrated in Fig1B,}  although fit to a slightly different equation,
$\textrm{\% Phase Separated} = 100 \times\left(\frac{1}{1+e^{-(P-P_c)/B}}\right)$.
\textcolor{black}{As with previous studies varying only temperature~\citep{Veatch08}, we find that the fraction of phase separated vesicles present in a population of vesicles at a given temperature/pressure does not depend on the order in which temperature or pressures are sampled, suggesting the transition is fully reversible.  }
\linebreak
\linebreak
\noindent \textit{Tadpole loss of righting reflex (LRR) measurements:}  
Studies with \textit{Xenopus laevis} tadpoles were conducted in compliance with the US Department of Health and Human Services Guide for the Care and Use of Laboratory Animals and were approved by the University of Michigan IACUC. \textit{Xenopus laevis} embryos were collected, fertilized, and dejellied as described previously~\cite{Miller09}. Embryos were stored at room temperature in 0.1X MMR (1X MMR = 100mM NaCl, 2mM KCl, 2mM CaCl$_2$, 1mM MgCl$_2$, 5mM Hepes, pH 7.4) and allowed to develop to swimming tadpole stage (stage 43-45).  

The LRR response was determined by placing tadpoles in a clean glass container filled with 50ml well water and the specified concentration of n-alcohols.  At 10min intervals, tadpoles were probed with a smooth glass rod and their responses  recorded.  Measurements with ethanol or ethanol and hexadecanol in equilibrated solutions were conducted on three separate occasions each with five tadpoles per condition, totaling fifteen tadpoles per condition.  Some measurements were conducted double-blind in order to avoid possible systematic bias in the behavioral scoring. Fewer tadpoles were probed for the other $n$-alcohol combinations investigated.  Error bars on LRR measurements are 68\% confidence intervals with binomial errors calculated according to:
$LB = 1-BetaInv(0.32/2, n-nk, nk+1) $ and 
 $UB = 1-BetaInv(1-0.32/2, n-nk+1, nk)$ 
where $LB$ and $UB$ are the lower and upper bounds of the confidence interval, $n$ is the number of tadpoles investigated, $k$ is the fractional LRR, and BetaInv is the Beta inverse cumulative distribution function.

To estimate $AC_{50}$, LRR measurements at a range of ethanol concentrations were fit to the form $\textrm{LRR} = 1- \frac{1}{1+e^{-([EtOH]-AC50)/B}}$.  Errors reported are 68\% confidence intervals of parametric uncertainty in $AC_{50}$.  These are calculated by inverting the expectation value of the Fisher Information and then taking the square root of its diagonal entry.  Error bars are much larger for hexadecanol containing titrations because all data is taken below the extrapolated $AC_{50}$ concentration. 
\linebreak
\linebreak
\noindent \textbf{RESULTS}

Fig. 2A shows that incubating super-saturated solutions of hexadecanol (n=16 alcohol) with isolated Rat Basophilic Leukemia (RBL) and XtC-2 derived GPMVs acts to raise $T_c$.  \textcolor{black}{Here, GPMVs derived from both mammalian (RBL) and \textit{Xenopus laevis} (XtC-2) cells were used to demonstrate robustness of this effect across species, since later measurements are conducted in  \textit{Xenopus laevis} tadpoles.  } The sign change in $\Delta T_c$ for hexadecanol compared to shorter chain n-alcohols mirrors past work which found that hexadecanol acted to increase chain order within native isolated membranes, while shorter anesthetic n-alcohols acted to decrease chain order.~\citep{Miller89}. We find that the effect on $T_c$ is approximately additive when ethanol (n=2 alcohol) and hexadecanol are used in combination, with $3\mu$M hexadecanol required to counter the effect of an $AC_{50}$ of ethanol (120mM) over a range of ethanol and hexadecanol concentrations (Fig. 2A,B). 

\begingroup
    \includegraphics[width=3.75in]{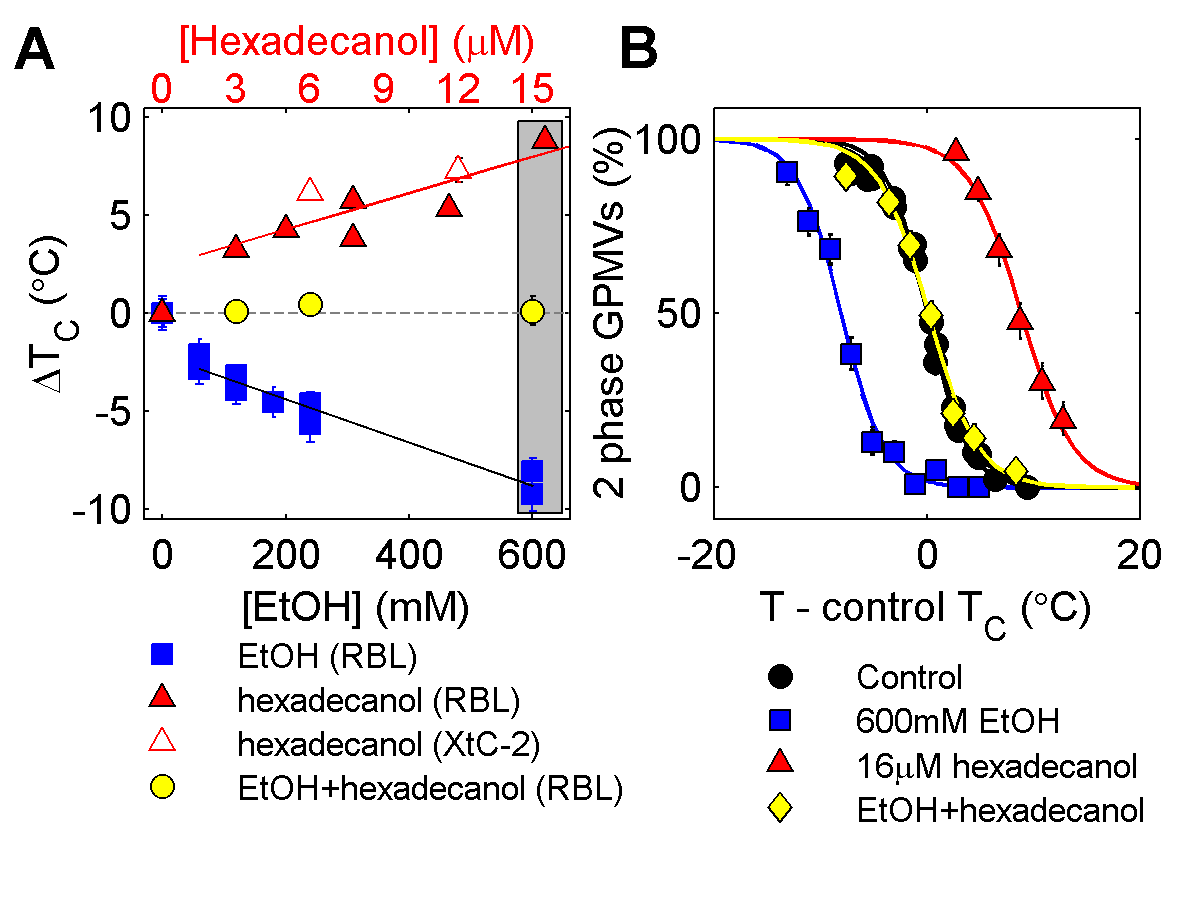}
{\normalsize
\linebreak
\noindent {\bf Fig. 2.} Hexadecanol raises $T_c$ in GPMVs from rat (RBL) and \textit{Xenopus} (XtC-2) cell lines and can counteract the $T_c$ lowering effects of ethanol. (A) Values  indicate the average shift in $T_c$  ($\Delta T_c$) in a population of vesicles upon treatment with the compounds indicated. Solutions containing hexadecanol were prepared to be super-saturated as described in Methods.  Each point represents a single measurement and error bounds represent the 68\% confidence interval on the extrapolated $\Delta T_c$.   (B) Plots showing fraction of phase separated vesicles vs. temperature for the three points inside the gray box in A. 
\linebreak
\linebreak
}
\endgroup
\linebreak

\begingroup

	\includegraphics[width=6.75in]{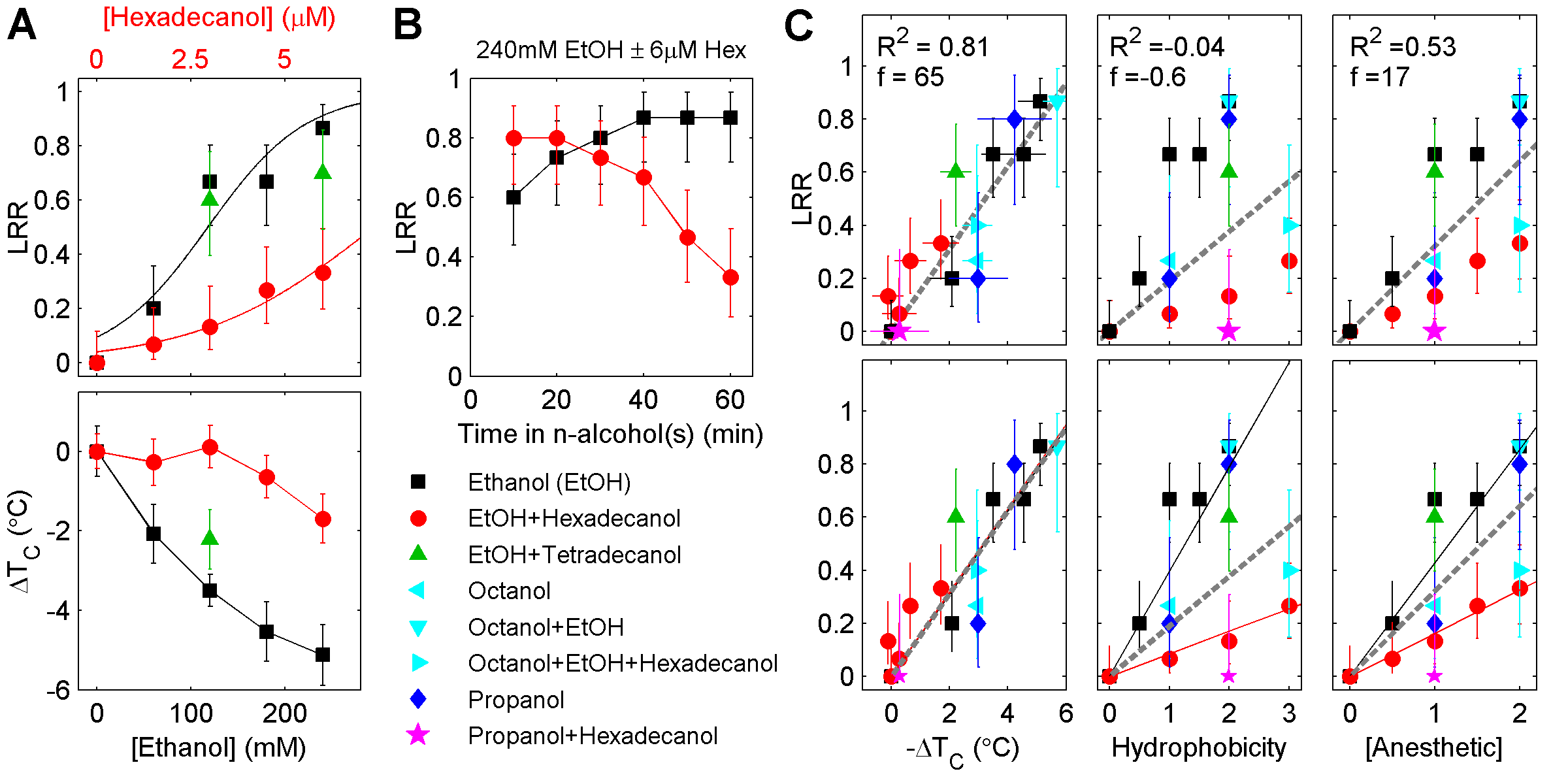}
\normalsize{    
\noindent {\bf Fig. 3.} (A) (Upper panel) Tadpole loss of righting reflex (LRR) for a titration of ethanol alone and combinations of ethanol and hexadecanol (EtOH+Hex) or ethanol and tetradecanol (EtOH+Tet) measured after 1h incubation in equilibrated solutions. At a given ethanol concentration,the presence of hexadecanol increases the fraction of tadpoles which respond to stimulus.  (Lower panel)  $\Delta T_c$ in RBL derived GPMVs for identical titrations of ethanol and EtOH+Hex. All solutions contain the ethanol concentration indicated by the lower horizontal axis. Red circle points additionally contain  hexadecanol concentrations indicated by the upper horizontal axis, and green triangle points additionally contain either 5 or 10$\mu$M tetradecanol.  (B) Time-course of LRR for one ethanol and ethanol+hexadecanol combination.  \textcolor{black}{ (C) (Left) Points in A replotted as LRR vs $\Delta T_c$, including additional experiments with other n-alcohol combinations as indicated in the legend. (Center) LRR  plotted vs. aggregate hydrophobicity, tabulated by summing the concentration of each n-alcohol present normalized by its $AC_{50}$~\citep{Pringle81}, using $3\mu$M as a proxy $AC_{50}$ for hexadecanol and $5\mu$M as a proxy $AC_{50}$ for tetradecanol. (Right) LRR  plotted vs. the net anesthetic concentration, tabulated by summing the concentration of each n$<$14 alcohol anesthetic present normalized by its $AC_{50}$~\citep{Pringle81}. (Top panels) In each case the dashed line represents a best fit to the data with R$^2$ values giving the fraction of explained variance and f values comparing the quality of the fit to a null model where all points have the same probability.  All fits are linear and constrained to go through the origin.  (Bottom Panels) In each case the black line is fit to all conditions that exclude hexadecanol, the red line is fit to all conditions that include hexadecanol, and the gray dashed line is fit to all points as in the upper panels.  These fits are substantially different from each other except when plotted vs $\Delta T_c$, indicating that $\Delta T_c$ has the most quantitative predictive power across chemical species.}
\linebreak
\linebreak
}

\endgroup

\textcolor{black}{ Mimicking results from our previous study with n-alcohols~\citep{Gray13}, we again find that relatively small concentrations of both hexadecanol and ethanol  lead to a steep change in $T_c$, while larger concentrations lead to a less steep and approximately linear regime. This nonlinearity implies an interaction between the added molecules, but we cannot specify the nature of their interaction from our measurements.  It could be that the concentration of n-alcohol in the membrane becomes a sub-linear function of their concentration in the bulk.  Past work has demonstrated that n-alcohol membrane partition coefficients can depend on n-alcohol concentration~\citep{Fraser1991}, membrane composition~\citep{Fraser1991,Trandum00}, and temperature~\citep{Trandum00}, and it is possible that these or related effects could lead to the observed nonlinearity.  Another possibility is that the addition of small molecules could increase the area of the membrane, thereby reducing its tension.  This could have a direct effect on T$_c$~\citep{Portet12}, or could alter the membrane's affinity for additional small molecules.  It is also possible that membrane concentration is linear with that of the bulk, but that $T_c$ is a nonlinear function of  n-alcohol concentration in the membrane. We note that ethanol can induce a novel interdigitated membrane phase in purified phosphatidylcholine membranes when at high enough concentration~\citep{Rowe90,McIntosh01} so it is plausible that specific interactions between certain lipids and alcohols could drive at least some of the nonlinearity that we observe.  At $AC_{50}$, small molecules make up several mol\% of the membrane~\citep{Pringle81}, a relatively high concentration which makes it plausible that n-alcohols interact directly, or indirectly through membrane mediated mechanisms that can be very sensitive near the critical point~\citep{Widom67}. }

Our observation that hexadecanol can counteract the effects of ethanol in GPMVs led us to speculate that hexadecanol could antagonize ethanol anesthesia.  To test this hypothesis, we examined the tadpole loss of righting reflex (LRR) in \textit{Xenopus laevis} tadpoles (Fig.3A). We also measured the extent to which identical n-alcohol treatments alter $T_c$ in RBL derived GPMVs. Here equilibrated solutions of hexadecanol and ethanol were used, leading to small differences from the results presented in Fig 2. 
In ethanol alone, 50\% of tadpoles did not exhibit a righting reflex upon inverting with a glass rod at 118$\pm15$mM ethanol in agreement with previous reports of 120mM~\citep{Pringle81}. Tadpoles incubated in ethanol and hexadecanol remained alert to much higher ethanol concentrations, more than doubling the extrapolated $AC_{50}$ value to 294$\pm$61mM. Co-incubation with tetradecanol (n=14) had little effect on either GPMV transition temperatures or on the tadpole LRR, suggesting that competitive inhibition at a putative ethanol binding site is not the mechanism through which hexadecanol reverses the effects of ethanol on LRR.

Fig.3B shows the time dependent effects of ethanol and ethanol-hexadecanol mixtures on tadpole LRR.  LRR  is not initially affected by the presence of hexadecanol, but instead reduces LRR over the span of 1h compared to those incubated in ethanol alone. Experiments were not extended beyond 1h as we observed some adaptation of ethanol treated tadpoles beyond this time-frame. The slow onset of hexadecanol action is consistent with past work using radio-labeled n-alcohols, which demonstrated that their absorption dynamics depend on carbon length, with longer n-alcohols requiring longer times for incorporation~\citep{Miller89}. \textcolor{black}{We note that the apparent difference in LRR for ethanol+hexadecanol vs. ethanol alone is not significant at early time-points, nor is the apparent upward trend in LRR vs. time for tadpoles treated with ethanol alone. }  In Fig.3C we plot LRR vs $\Delta T_c$, aggregate hydrophobicity, or anesthetic concentration for a range of conditions.  \textcolor{black}{We find that $\Delta T_c$ captures substantially more of the variance in LRR across trials (R$^2=.82$) than summed anesthetic potency (R$^2=.53$), while aggregate hydrophobicity has no predictive power ($R<0$).  This is partially due to our carefully chosen treatments, many of which are hydrophobic but raise critical temperatures and antagonize LRR. Furthermore, we find that $\Delta T_c$ predicts LRR with the same function whether for treatments containing hexadecanol or treatments without hexadecanol, while both aggregate hydrophobicity and summed anesthetic potency do not.  Together this demonstrates that $\Delta T_c$ predicts LRR across molecular species.}

This indicates that $\Delta T_c$ is significantly more predictive of anesthetic potency than aggregate hydrophobicity or anesthetic concentration for the n-alcohol mixtures examined. 

The extremely low solubility of hexadecanol presents some experimental difficulties. Ethanol is required as a co-solvent in these measurements and this prevented a more systematic exploration of how hexadecanol modulates anesthesia mediated by other n-alcohols and anesthetics.  Our reliance on ethanol as a general anesthetic introduces technical 
challenges since it has low potency, can harbor impurities that modulate LRR measurements~\citep{Miller89}, and produces alternate functional outcomes when used at high concentrations~\citep{Downes96}.  While a more water soluble $T_c$ raising compound would alleviate some of these problems, most membrane soluble molecules with reasonable water solubility either lower GPMV transition temperatures or leave them unchanged.   Notable exceptions are some detergents~\citep{Zhou13}, which are not suitable for this investigation because they also permeabilize membranes.  Molecules which raise $T_c$ must partition more strongly than the average component into one low temperature membrane phase~\citep{Meerschaert15} and our observations suggest that this often requires a large hydrophobic interaction area, with consequent low solubility in water.
\linebreak

\begingroup
    \includegraphics[width=3.75in]{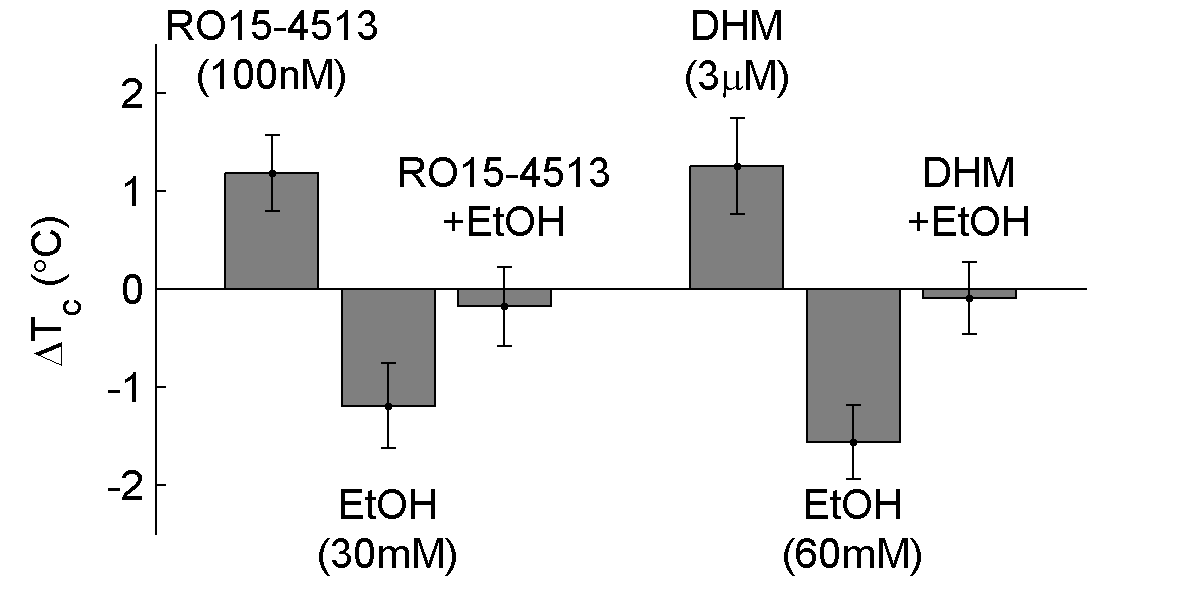}

{\normalsize
\noindent {\bf Fig. 4.} Ro15-4513 and DHM block the acute toxicity and intoxicating effect of ethanol. Each raise $T_c$ and cancel the effects of ethanol when added to GPMVs at the same concentration at which they are effective \textit{in vivo}. 
\linebreak
\linebreak
}
\endgroup

Several other compounds are reported to reverse the intoxicating effects of ethanol in both cultured neurons and intact organisms, and their effects on RBL derived GPMVs are shown in Fig 4. $100$nM RO15-4513, a therapeutic agent used to reverse acute ethanol toxicity, has been demonstrated to reverse effects of $30$mM ethanol in the GABA current in oocytes~\citep{Wallner06}. We find that $100$nM RO15-4513 raises critical temperatures in GPMVs by $1.2 \pm 0.4 ^\circ$C.  Similarly, $3\mu$M Dihydromyricetin (DHM) was recently reported to antagonize the effects of $60$mM ethanol on the GABA-induced current in acutely dissociated rat hippocampal slices as well as on behavioral assays \textit{in vivo}~\citep{Shen12}.  We find that $3\mu$M DHM raises critical temperatures by $1.3 \pm 0.5 ^\circ$C. In each of these cases, the effect of these compounds on $T_c$ appears to saturate at concentrations well below their solubility.  While we have no mechanistic explanation for this saturation, we note that it is mirrored in past behavioral measurements~\citep{Wallner06,Shen12} \textcolor{black}{which find that they are only able to counter ethanol effects to concentrations well below its AC$_{50}$ of 120mM. }

While the correlation between $\Delta T_c$ and anesthetic function is robust for the compounds investigated,  we note that an additional compound, menthol, raises $T_c$ by nearly 2$^\circ$C when added to RBL derived GPMVs at 100$\mu$M~\citep{Raghunathan15}, yet acts as an anesthetic in tadpole LRR measurements at somewhat lower concentration (AC$_{50}\approx$20$\mu$M)~\citep{Watt08}. \textcolor{black}{This could be due to menthol having specific interactions with GABA$_A$ receptors, as has been proposed previously~\citep{Watt08}}. It is also possible that menthol has different effects on \textit{Xenopus laevis} neuronal vs. mammalian immune plasma membranes.
\linebreak

\begingroup
    \includegraphics[width=3.75in]{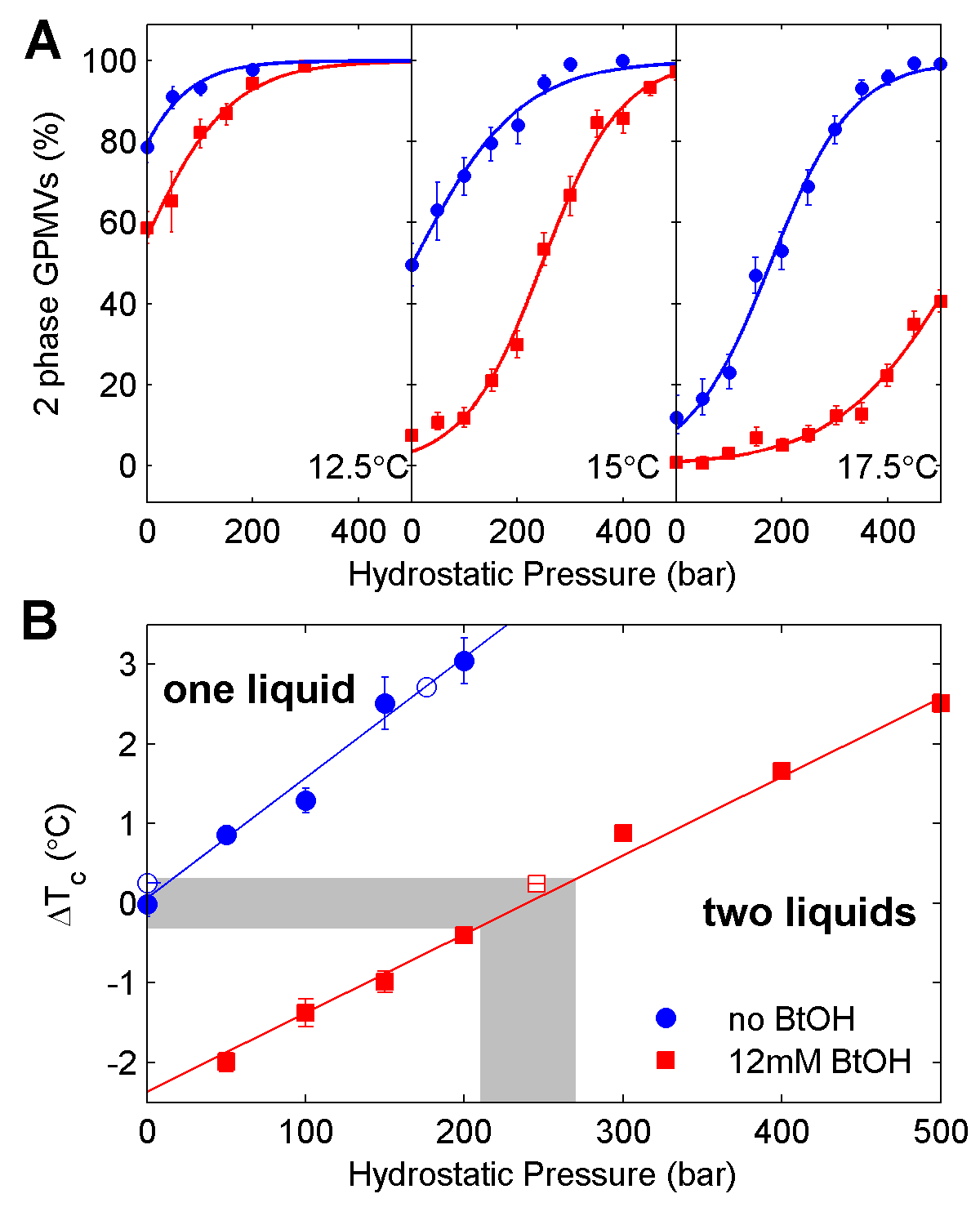}

{\normalsize
\noindent {\bf Fig. 5.} (A) The fraction of vesicles which are macroscopically phase separated is plotted as a function of hydrostatic pressure at three different temperatures, both for control vesicles and vesicles incubated with 12mM bBtOH.  In each case, increasing the pressure leads to an increase in the fraction of vesicles which are macroscopically phase separated.  (B)  $T_c$ is raised with increasing hydrostatic pressure in both control GPMVs and GPMVs incubated in butanol (BtOH). Here, 240$\pm$30 bar of hydrostatic pressure is required to reverse the effects of 12mM BtOH (shaded region). Closed symbols are obtained by extrapolating to find $T_c$ from data acquired at constant pressure while open symbols are obtained by extrapolating to find $P_c$ at constant temperature.  At temperatures above $T_c$ most vesicles are composed of a macroscopically uniform single liquid, while below $T_c$ most are separated into two co-existing liquid phases.
\linebreak
\linebreak
}
\endgroup
\linebreak
\linebreak
\noindent \textbf{DISCUSSION}

We also investigated the effects of hydrostatic pressure on GPMVs in the presence and absence of butanol (n=4 alcohol).  It has long been known that 150-200 bar of pressure reverses animal anesthesia~\citep{Johnson50}.  Here we show that the fraction of vesicles with two phase coexistence increases monotonically with pressure, for vesicles incubated with and without butanol (Fig 5A).   We observe that 240$\pm$30 bar is needed to counter the effects of one $AC_{50}$ of butanol (12mM~\citep{Pringle81}) on the $T_c$ of RBL derived GPMVs (Fig 5B).    Pressure reversal of general anesthesia is not accounted for by current models involving direct binding of anesthetics to channels because protein conformational equilibria are typically sensitive only to much larger pressures ($>$1000 bar)~\citep{Moss91,Mozhaev96}. These smaller pressures can have large effects on some membrane properties, and past work has shown that 200 bar raises the main chain melting temperature in synthetic single component lipid systems by $\sim4.5^\circ$C ~\citep{Liu77,WInter09}, even more than is observed here for the miscibility transition in GPMVs (2-3$^\circ$C).

While our results are not sufficient to specify a mechanism, they do strongly suggest that \textcolor{black}{n-alcohol} anesthetic inhibition of membrane heterogeneity plays an important role in mediating anesthesia, likely by interfering with normal membrane regulation of ion channels. \textcolor{black}{We have previously argued that the presence of a critical point in GPMVs near or below room temperature leads to extended and dynamic composition fluctuations at growth temperature (typically 37$^\circ$C), both in isolated vesicles and in the intact plasma membranes from which they are derived~\citep{Veatch08,Machta11}.  Theory and experiment suggest that the size of fluctuations varies as $(T-T_c)^{-1}$~\citep{Veatch08, Zhao13}, so that lowering $T_c$ through the addition of an n-alcohol anesthetic is expected to reduce the size of compositional heterogeneity at fixed growth temperature (Fig. 6). This suggests several plausible mechanisms through which changes in plasma membrane $T_c$ could modulate channel function. First, reducing the size of membrane structures could in principle increase (or decrease) accessibility of native regulators of channel function such as neurosteroids, phosphoinositides and G-proteins~\citep{Allen07}.  Second, different membrane domains are expected to have differing physical properties such as hydrophobic thickness, lateral pressure profiles, viscosity, and bending rigidity~\citep{Diaz-Rohrer14,Lin13,Niemela07,Cicuta07,Baumgart03}.  Destabilizing these membrane domains could act to modulate the response of channels directly by stabilizing different internal protein states. Thirdly, membrane domains are argued to play important roles in organizing channels at neuronal synapses~\citep{Allen07} and destabilizing these domains could lead to altered localization of channels with disruption of higher level neuronal functions.  Each of these mechanistic possibilities have distinct experimental signatures that can be explored in future studies.    }
\linebreak

\begingroup
    \includegraphics[width=3.75in]{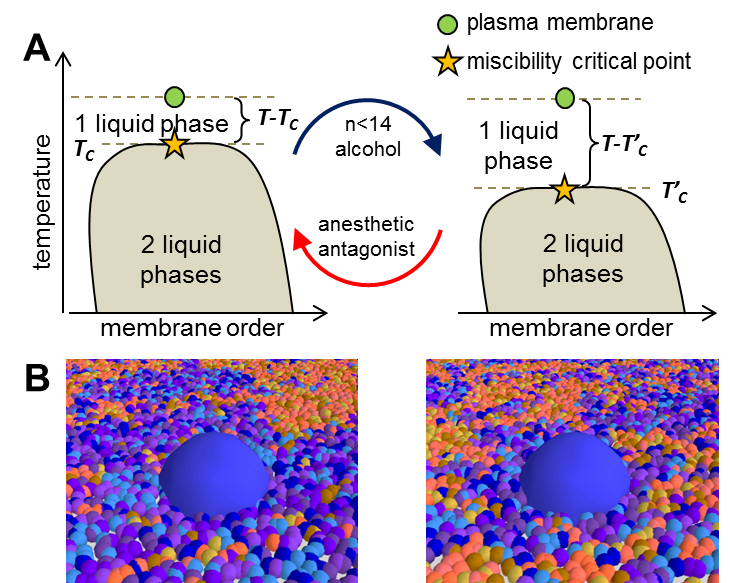}

{\normalsize
\textcolor{black}{
\noindent {\bf Fig. 6.} Anesthetics lower the transition temperature of a biologically tuned critical point which could lead to mis-regulation of ion channels and other membrane bound proteins.  (A) Schematic phase diagrams for the plasma membrane of an untreated and n-alcohol treated cell.  Guided by experiments~\citep{Veatch08} we hypothesize that the plasma membrane lies in the white region where lipids, proteins and other membrane components are well mixed macroscopically into a single two dimensional liquid phase.  However, due to their close proximity to the critical point (star), thermal fluctuations lead to relatively large domains enriched in particular components.  When cooled into the gray two phase region, GPMVs separate into two coexisting liquid phases termed liquid-ordered and liquid-disordered. Several n-alcohol general anesthetics lower the critical temperature of the membrane~\citep{Gray13}, changing the distance above the critical point, $T-T_c$. Here we also show that treatments which antagonize anesthetic action \textit{raise} critical temperatures reversing the effects of n-alcohols on $T_c$.  (B) Under normal conditions a hypothetical ion channel (large blue inclusion) has a tendency to inhabit relatively large domains enriched in particular lipids and proteins.  When the membrane is taken away from the critical point,  the structure of these domains is altered, possibly leading to changes in ion channel gating and function through a variety of mechanisms discussed in the text.
}
\linebreak
\linebreak
}

\endgroup

 \textcolor{black}{One naive prediction of our model is not borne out by experiment, namely that the functional effects of anesthetics are not readily reversed by lowering ambient temperature~\citep{Franks82}, even though this should reverse anesthetic effects on the parameter $T-T_c$. While this finding is at odds with a simple interpretation of our results, we note that other biological functions have been shown to be explicitly temperature compensated~\citep{Oleksiuk11,Kidd15}. In these examples individual reaction rates often display sharp temperature sensitivity, but the effects of these changes on certain important axes (e.g. the timing of the circadian clock) roughly cancel, leaving function robust~\citep{Daniels08}.  
Thus, while many cellular processes relevant to neuronal function have sensitive temperature dependence~\citep{Smith08, Thompson85,Tang10}, we expect that organisms, especially those able to live over a range of temperatures, have evolved to a regime where neural function is temperature compensated~\citep{Tang10} and is thus surprisingly insensitive to changes in temperature. However, we might still expect that triggering only temperature's effects on membrane criticality in an uncompensated way, as we argue occurs with n-alcohol anesthesia, might lead to dramatic changes in function~\citep{Daniels08}. We note that past studies which demonstrate a lack of temperature dependence on anesthetic potency have conducted behavioral measurements in temperature adapted animals~\citep{Meyer01,Cherkin64}, where other studies have found evidence for concurrent changes in membrane composition~\citep{Baranska69,Anderson70}.  Acute temperature changes mimicking or exceeding those produced by relevant concentrations of anesthetic ($\pm$4$^\circ$C) do produce behavioral changes in animals ranging from lethargy to death~\citep{Meyer01}. }

Finally, while the evidence for anesthetic-channel interactions is significant, several features may suggest the interaction is more akin to two dimensional solvent-solute rather than binding site-ligand interactions. Some channels contain multiple proposed binding sites for low potency anesthetics, each with low affinity but diffusion limited association rates~\cite{Xu00}, with clinically relevant concentrations of anesthetic as high as several mol\% of the membrane~\citep{Janoff81}. Additionally, many anesthetic sensitive channels are also sensitive to membrane properties, and in particular cholesterol modulation both in reconstituted~\citep{Bristow87} and cellular systems~\citep{Sooksawate01b}, suggesting a commonality with broader regulation by membrane domains.  
 Our results suggest that these effects on channel function are likely to be altered by anesthetics through their effects on membrane mixing even without binding of anesthetic to protein targets.
\linebreak
\linebreak
\noindent \textbf{CONCLUSION}

Overall, the results presented here demonstrate that a condition's effect on GPMV $T_c$ is more predictive of anesthetic potency than hydrophobicity.  By exploiting rare conditions where $\Delta T_c >0$ we have shown that behavioral measures and existing electrophysiological assays for anesthesia are remarkably tied to the membrane's thermodynamic propensity to form small domains. While our results do not suggest a specific mechanism through which these conditions reverse anesthesia, they do support the hypothesis that at least some n-alcohol targets may be influenced through effects on membrane criticality.   
\linebreak
\linebreak
\noindent \textbf{Acknowledgements} We thank Keith Miller and James Sethna for helpful converstaions, Ned Wingreen for a close reading of the manuscript and Margaret Burns, Jing Wu, and Kathleen Wisser for assistance with some experiments.  Research was funded by the NIH (R01 GM110052 to SLV and R01 GM112794 to ALM), startup funds from the University of Michigan (to SLV), EPSRC Programme grant (EP/J017566/1 to NJB), a EPSRC Centre for Doctoral Training Studentship awarded by the Institute of Chemical Biology to NLCM (EP/F500076/1) a Lewis-Sigler Fellowship (BBM) and by NSF PHY 0957573 (BBM).









\begin{thebibliography}{70}
\providecommand{\url}[1]{\texttt{#1}}
\providecommand{\urlprefix}{ }

\bibitem[Meyer(1899)]{Meyer99}
Meyer, H., 1899.
\newblock Zur Theorie der Alkoholnarkose.
\newblock \emph{Archiv für experimentelle Pathologie und Pharmakologie}
  42:109--118.

\bibitem[Gruner and Shyamsunder(1991)]{Gruner91}
Gruner, S.~M., and E.~Shyamsunder, 1991.
\newblock Is the mechanism of general anesthesia related to lipid membrane
  spontaneous curvature?
\newblock \emph{Ann N Y Acad Sci} 625:685--97.

\bibitem[Wodzinska et~al.(2009)Wodzinska, Blicher, and Heimburg]{Wodzinska09}
Wodzinska, K., A.~Blicher, and T.~Heimburg, 2009.
\newblock The thermodynamics of lipid ion channel formation in the absence and
  presence of anesthetics. BLM experiments and simulations.
\newblock \emph{Soft Matter} 5:3319--3330.

\bibitem[Ing{\'o}lfsson and Andersen(2011)]{Ingolfsson11}
Ing{\'o}lfsson, H.~I., and O.~S. Andersen, 2011.
\newblock Alcohol's Effects on Lipid Bilayer Properties.
\newblock \emph{Biophysical Journal} 101:847--855.

\bibitem[Herold et~al.(2014)Herold, Sanford, Lee, Schultz, Ing�lfsson,
  Andersen, and Hemmings]{Herold14}
Herold, K.~F., R.~L. Sanford, W.~Lee, M.~F. Schultz, H.~I. Ing�lfsson, O.~S.
  Andersen, and H.~C. Hemmings, 2014.
\newblock Volatile anesthetics inhibit sodium channels without altering bulk
  lipid bilayer properties.
\newblock \emph{The Journal of General Physiology} 144:545--560.

\bibitem[Franks and Lieb(1986)]{Franks86}
Franks, N.~P., and W.~R. Lieb, 1986.
\newblock Partitioning of long-chain alcohols into lipid bilayers: implications
  for mechanisms of general anesthesia.
\newblock \emph{Proc Natl Acad Sci U S A} 83:5116--20.

\bibitem[Franks and Lieb(1994)]{Franks94}
Franks, N.~P., and W.~R. Lieb, 1994.
\newblock Molecular and cellular mechanisms of general anaesthesia.
\newblock \emph{Nature} 367:607--14.

\bibitem[Mihic et~al.(1997)Mihic, Ye, Wick, Koltchine, Krasowski, Finn, Mascia,
  Valenzuela, Hanson, Greenblatt, Harris, and Harrison]{Mihic97}
Mihic, S.~J., Q.~Ye, M.~J. Wick, V.~V. Koltchine, M.~A. Krasowski, S.~E. Finn,
  M.~P. Mascia, C.~F. Valenzuela, K.~K. Hanson, E.~P. Greenblatt, R.~A. Harris,
  and N.~L. Harrison, 1997.
\newblock Sites of alcohol and volatile anaesthetic action on GABA(A) and
  glycine receptors.
\newblock \emph{Nature} 389:385--389.

\bibitem[Borghese et~al.(2006)Borghese, Werner, Topf, Baron, Henderson, Boehm,
  Blednov, Saad, Dai, Pearce, Harris, Homanics, and Harrison]{Borghese06}
Borghese, C.~M., D.~F. Werner, N.~Topf, N.~V. Baron, L.~A. Henderson, S.~L.
  Boehm, Y.~A. Blednov, A.~Saad, S.~Dai, R.~A. Pearce, R.~A. Harris, G.~E.
  Homanics, and N.~L. Harrison, 2006.
\newblock An Isoflurane- and Alcohol-Insensitive Mutant GABAA Receptor ?1
  Subunit with Near-Normal Apparent Affinity for GABA: Characterization in
  Heterologous Systems and Production of Knockin Mice.
\newblock \emph{Journal of Pharmacology and Experimental Therapeutics}
  319:208--218.

\bibitem[Nury et~al.(2011)Nury, Van~Renterghem, Weng, Tran, Baaden, Dufresne,
  Changeux, Sonner, Delarue, and Corringer]{Nury11}
Nury, H., C.~Van~Renterghem, Y.~Weng, A.~Tran, M.~Baaden, V.~Dufresne, J.~P.
  Changeux, J.~M. Sonner, M.~Delarue, and P.~J. Corringer, 2011.
\newblock X-ray structures of general anaesthetics bound to a pentameric
  ligand-gated ion channel.
\newblock \emph{Nature} 469:428--+.

\bibitem[Forman and Miller(2016)]{Forman16}
Forman, S.~A., and K.~W. Miller, 2016.
\newblock Mapping General Anesthetic Sites in Heteromeric $\gamma$-Aminobutyric
  Acid Type A Receptors Reveals a Potential For Targeting Receptor Subtypes.
\newblock \emph{Anesth Analg.} in press.

\bibitem[Xu et~al.(2000)Xu, Seto, Tang, and Firestone]{Xu00}
Xu, Y., T.~Seto, P.~Tang, and L.~Firestone, 2000.
\newblock NMR study of volatile anesthetic binding to nicotinic acetylcholine
  receptors.
\newblock \emph{Biophys J} 78:746--51.

\bibitem[Brannigan et~al.(2010)Brannigan, LeBard, Hénin, Eckenhoff, and
  Klein]{Brannigan09}
Brannigan, G., D.~N. LeBard, J.~Hénin, R.~G. Eckenhoff, and M.~L. Klein, 2010.
\newblock Multiple binding sites for the general anesthetic isoflurane
  identified in the nicotinic acetylcholine receptor transmembrane domain.
\newblock \emph{Proceedings of the National Academy of Sciences}
  107:14122--14127.

\bibitem[Simons and Ikonen(1997)]{Simons97}
Simons, K., and E.~Ikonen, 1997.
\newblock Functional rafts in cell membranes.
\newblock \emph{Nature} 387:569--72.

\bibitem[Anderson and Jacobson(2002)]{Anderson02}
Anderson, R.~G., and K.~Jacobson, 2002.
\newblock A role for lipid shells in targeting proteins to caveolae, rafts, and
  other lipid domains.
\newblock \emph{Science} 296:1821--5.

\bibitem[Allen et~al.(2007)Allen, Halverson-Tamboli, and Rasenick]{Allen07}
Allen, J.~A., R.~A. Halverson-Tamboli, and M.~M. Rasenick, 2007.
\newblock Lipid raft microdomains and neurotransmitter signalling.
\newblock \emph{Nat Rev Neurosci} 8:128--140.

\bibitem[Veatch and Keller(2005)]{Veatch05}
Veatch, S.~L., and S.~L. Keller, 2005.
\newblock Seeing spots: Complex phase behavior in simple membranes.
\newblock \emph{Biochimica et Biophysica Acta (BBA) - Molecular Cell Research}
  1746:172 -- 185.

\bibitem[Baumgart et~al.(2007)Baumgart, Hammond, Sengupta, Hess, Holowka,
  Baird, and Webb]{Baumgart07}
Baumgart, T., A.~T. Hammond, P.~Sengupta, S.~T. Hess, D.~A. Holowka, B.~A.
  Baird, and W.~W. Webb, 2007.
\newblock Large-scale fluid/fluid phase separation of proteins and lipids in
  giant plasma membrane vesicles.
\newblock \emph{Proc Natl Acad Sci U S A} 104:3165--70.

\bibitem[Veatch et~al.(2008)Veatch, Cicuta, Sengupta, Honerkamp-Smith, Holowka,
  and Baird]{Veatch08}
Veatch, S.~L., P.~Cicuta, P.~Sengupta, A.~Honerkamp-Smith, D.~Holowka, and
  B.~Baird, 2008.
\newblock Critical fluctuations in plasma membrane vesicles.
\newblock \emph{Acs Chemical Biology} 3:287--293.

\bibitem[Honerkamp-Smith et~al.(2008)Honerkamp-Smith, Cicuta, Collins, Veatch,
  den Niljs, Schick, and Keller]{HonerkampSmith08}
Honerkamp-Smith, A.~R., P.~Cicuta, M.~D. Collins, S.~L. Veatch, .~den Niljs,
  M.~Schick, and S.~L. Keller, 2008.
\newblock Line Tensions, Correlation Lengths, and Critical Exponents in Lipid
  Membranes Near Critical Points.
\newblock \emph{Biophysical Journal} 95:236--246.

\bibitem[Gray et~al.(2013)Gray, Karslake, Machta, and Veatch]{Gray13}
Gray, E., J.~Karslake, B.~B. Machta, and S.~L. Veatch, 2013.
\newblock Liquid general anesthetics lower critical temperatures in plasma
  membrane vesicles.
\newblock \emph{Biophys J} 105:2751--9.

\bibitem[Pringle et~al.(1981)Pringle, Brown, and Miller]{Pringle81}
Pringle, M.~J., K.~B. Brown, and K.~W. Miller, 1981.
\newblock Can the lipid theories of anesthesia account for the cutoff in
  anesthetic potency in homologous series of alcohols?
\newblock \emph{Mol Pharmacol} 19:49--55.

\bibitem[Barsumian et~al.(1981)Barsumian, Isersky, Petrino, and
  Siraganian]{Barsumian81}
Barsumian, E.~L., C.~Isersky, M.~G. Petrino, and R.~P. Siraganian, 1981.
\newblock IgE-induced histamine release from rat basophilic leukemia cell
  lines: isolation of releasing and nonreleasing clones.
\newblock \emph{European Journal of Immunology} 11:317--323.

\bibitem[Pudney et~al.(1973)Pudney, Varma, and Leake]{Pudney73}
Pudney, M., M.~G. Varma, and C.~J. Leake, 1973.
\newblock Establishment of a cell line (XTC-2) from the South African clawed
  toad, Xenopus laevis.
\newblock \emph{Experientia} 29:466--7.

\bibitem[Fridriksson et~al.(1999)Fridriksson, Shipkova, Sheets, Holowka, Baird,
  and McLafferty]{Fridriksson99}
Fridriksson, E.~K., P.~A. Shipkova, E.~D. Sheets, D.~Holowka, B.~Baird, and
  F.~W. McLafferty, 1999.
\newblock Quantitative Analysis of Phospholipids in Functionally Important
  Membrane Domains from RBL-2H3 Mast Cells Using Tandem High-Resolution Mass
  Spectrometry.
\newblock \emph{Biochemistry} 38:8056--8063.

\bibitem[Levental et~al.(2016)Levental, Lorent, Lin, ad~Michal A.~Surma,
  Stockenbojer, Gorfe, and Levental]{Levental16}
Levental, K.~R., J.~H. Lorent, X.~Lin, A.~D.~S. ad~Michal A.~Surma, E.~A.
  Stockenbojer, A.~A. Gorfe, and I.~Levental, 2016.
\newblock Polyunsaturated Lipids Regulate Membrane Domain Stability by Tuning
  Membrane Order.
\newblock \emph{Biophysical Journal} 110:1800--1810.

\bibitem[McCarthy et~al.(2015)McCarthy, Ces, Law, Seddon, and
  Brooks]{McCarthy15}
McCarthy, N. L.~C., O.~Ces, R.~V. Law, J.~M. Seddon, and N.~J. Brooks, 2015.
\newblock Separation of liquid domains in model membranes induced with high
  hydrostatic pressure.
\newblock \emph{Chem. Commun.} 51:8675--8678.

\bibitem[Purushothaman et~al.(2015)Purushothaman, Cicuta, Ces, and
  Brooks]{Purushothaman15}
Purushothaman, S., P.~Cicuta, O.~Ces, and N.~J. Brooks, 2015.
\newblock Influence of High Pressure on the Bending Rigidity of Model
  Membranes.
\newblock \emph{The Journal of Physical Chemistry B} 119:9805--9810.
  26146795.

\bibitem[Gray et~al.(2015)Gray, Diaz-Vazquez, and Veatch]{Gray15}
Gray, E.~M., G.~Diaz-Vazquez, and S.~L. Veatch, 2015.
\newblock Growth Conditions and Cell Cycle Phase Modulate Phase Transition
  Temperatures in RBL-2H3 Derived Plasma Membrane Vesicles.
\newblock \emph{PLoS ONE} 10:1--16.

\bibitem[Miller and Bement(2009)]{Miller09}
Miller, A.~L., and W.~M. Bement, 2009.
\newblock Regulation of cytokinesis by Rho GTPase flux.
\newblock \emph{Nat Cell Biol} 11:71--77.

\bibitem[Miller et~al.(1989)Miller, Firestone, Alifimoff, and
  Streicher]{Miller89}
Miller, K.~W., L.~L. Firestone, J.~K. Alifimoff, and P.~Streicher, 1989.
\newblock Nonanesthetic alcohols dissolve in synaptic membranes without
  perturbing their lipids.
\newblock \emph{Proc Natl Acad Sci U S A} 86:1084--7.

\bibitem[Fraser et~al.(1991)Fraser, Van~Gorkom, and Watts]{Fraser1991}
Fraser, D.~M., L.~C. Van~Gorkom, and A.~Watts, 1991.
\newblock Partitioning behaviour of 1-hexanol into lipid membranes as studied
  by deuterium NMR spectroscopy.
\newblock \emph{Biochimica et Biophysica Acta (BBA) - Biomembranes}
  1069:53--60.

\bibitem[Trandum et~al.(2000)Trandum, Westh, Jørgensen, and
  Mouritsen]{Trandum00}
Trandum, C., P.~Westh, K.~Jørgensen, and O.~G. Mouritsen, 2000.
\newblock A Thermodynamic Study of the Effects of Cholesterol on the
  Interaction between Liposomes and Ethanol.
\newblock \emph{Biophysical Journal} 78:2486 -- 2492.


\bibitem[Portet et~al.(2012)Portet, Gordon, and Keller]{Portet12}
Portet, T., S.~E. Gordon, and S.~L. Keller, 2012.
\newblock Increasing Membrane Tension Decreases Miscibility Temperatures; an
  Experimental Demonstration via Micropipette Aspiration.
\newblock \emph{Biophysical Journal} 103:L35 -- L37.

\bibitem[Rowe and A.(1990)]{Rowe90}
Rowe, E.~S., and C.~T. A., 1990.
\newblock Differential Scanning Calorimetric Studies of Ethanol Interactions
  with Distearoylphosphatidylcholine: Transition to the Interdigitated Phase.
\newblock \emph{Biochemistry} 29:10398--10404.

\bibitem[McIntosh et~al.(2001)McIntosh, Lin, Li, and hsien Huang]{McIntosh01}
McIntosh, T.~J., H.~Lin, S.~Li, and C.~hsien Huang, 2001.
\newblock The effect of ethanol on the phase transition temperature and the
  phase structure of monounsaturated phosphatidylcholines.
\newblock \emph{Biochimica et Biophysica Acta (BBA) - Biomembranes} 1510:219 --
  230.


\bibitem[Widom({1967})]{Widom67}
Widom, B., {1967}.
\newblock Plait points in 2 and 3-Component liquid mixtures.
\newblock \emph{{J. Chem. Phys.}} {46}:{3324}.

\bibitem[Downes and Courogen(1996)]{Downes96}
Downes, H., and P.~M. Courogen, 1996.
\newblock Contrasting effects of anesthetics in tadpole bioassays.
\newblock \emph{Journal of Pharmacology and Experimental Therapeutics}
  278:284--296.
\newblock
  \urlprefix\url{http://jpet.aspetjournals.org/content/278/1/284.abstract}.

\bibitem[Zhou et~al.(2013)Zhou, Maxwell, Sezgin, Lu, Liang, Hancock, Dial,
  Lichtenberger, and Levental]{Zhou13}
Zhou, Y., K.~N. Maxwell, E.~Sezgin, M.~Lu, H.~Liang, J.~F. Hancock, E.~J. Dial,
  L.~M. Lichtenberger, and I.~Levental, 2013.
\newblock Bile Acids Modulate Signaling by Functional Perturbation of Plasma
  Membrane Domains.
\newblock \emph{Journal of Biological Chemistry} 288:35660--35670.

\bibitem[Meerschaert and Kelly(2015)]{Meerschaert15}
Meerschaert, R.~L., and C.~V. Kelly, 2015.
\newblock Trace membrane additives affect lipid phases with distinct
  mechanisms: a modified Ising model.
\newblock \emph{Eur Biophys J} 44:227--33.

\bibitem[Wallner et~al.(2006)Wallner, Hanchar, and Olsen]{Wallner06}
Wallner, M., H.~J. Hanchar, and R.~W. Olsen, 2006.
\newblock Low-dose alcohol actions on $\alpha_4\beta_3\gamma$ GABA$_A$
  receptors are reversed by the behavioral alcohol antagonist Ro15-4513.
\newblock \emph{Proceedings of the National Academy of Sciences}
  103:8540--8545.

\bibitem[Shen et~al.(2012)Shen, Lindemeyer, Gonzalez, Shao, Spigelman, Olsen,
  and Liang]{Shen12}
Shen, Y., A.~K. Lindemeyer, C.~Gonzalez, X.~M. Shao, I.~Spigelman, R.~W. Olsen,
  and J.~Liang, 2012.
\newblock Dihydromyricetin As a Novel Anti-Alcohol Intoxication Medication.
\newblock \emph{The Journal of Neuroscience} 32:390--401.

\bibitem[Raghunathan et~al.(2015)Raghunathan, Ahsan, Ray, Nyati, and
  Veatch]{Raghunathan15}
Raghunathan, K., A.~Ahsan, D.~Ray, M.~K. Nyati, and S.~L. Veatch, 2015.
\newblock Membrane Transition Temperature Determines Cisplatin Response.
\newblock \emph{PLoS ONE} 10:e0140925.

\bibitem[Watt et~al.(2008)Watt, Betts, Kotey, Humbert, Griffith, Kelly,
  Veneskey, Gill, Rowan, Jenkins, and Hall]{Watt08}
Watt, E.~E., B.~A. Betts, F.~O. Kotey, D.~J. Humbert, T.~N. Griffith, E.~W.
  Kelly, K.~C. Veneskey, N.~Gill, K.~C. Rowan, A.~Jenkins, and A.~C. Hall,
  2008.
\newblock Menthol shares general anesthetic activity and sites of action on the
  GABAA receptor with the intravenous agent, propofol.
\newblock \emph{European Journal of Pharmacology} 590:120 -- 126.

\bibitem[Johnson and Flagler(1950)]{Johnson50}
Johnson, F.~H., and E.~A. Flagler, 1950.
\newblock Hydrostatic pressure reversal of narcosis in tadpoles.
\newblock \emph{Science} 112:91--2.

\bibitem[Moss et~al.(1991)Moss, Lieb, and Franks]{Moss91}
Moss, G.~W., W.~R. Lieb, and N.~P. Franks, 1991.
\newblock Anesthetic inhibition of firefly luciferase, a protein model for
  general anesthesia, does not exhibit pressure reversal.
\newblock \emph{Biophysical Journal} 60:1309--1314.

\bibitem[Mozhaev et~al.(1996)Mozhaev, Heremans, Frank, Masson, and
  Balny]{Mozhaev96}
Mozhaev, V.~V., K.~Heremans, J.~Frank, P.~Masson, and C.~Balny, 1996.
\newblock High pressure effects on protein structure and function.
\newblock \emph{Proteins} 24:81--91.

\bibitem[Liu and Kay(1977)]{Liu77}
Liu, N.-I., and R.~L. Kay, 1977.
\newblock Redetermination of the pressure dependence of the lipid bilayer phase
  transition.
\newblock \emph{Biochemistry} 16:3484--3486.

\bibitem[Winter and Jeworrek(2009)]{WInter09}
Winter, R., and C.~Jeworrek, 2009.
\newblock Effect of pressure on membranes.
\newblock \emph{Soft Matter} 5:3157--3173.

\bibitem[Machta et~al.(2011)Machta, Papanikolaou, Sethna, and Veatch]{Machta11}
Machta, B.~B., S.~Papanikolaou, J.~P. Sethna, and S.~L. Veatch, 2011.
\newblock Minimal model of plasma membrane heterogeneity requires coupling
  cortical actin to criticality.
\newblock \emph{Biophys J} 100:1668--77.

\bibitem[Zhao et~al.(2013)Zhao, Wu, and Veatch]{Zhao13}
Zhao, J., J.~Wu, and S.~L. Veatch, 2013.
\newblock Adhesion stabilizes robust lipid heterogeneity in supercritical
  membranes at physiological temperature.
\newblock \emph{Biophys J} 104:825--34.

\bibitem[Diaz-Rohrer et~al.(2014)Diaz-Rohrer, Levental, Simons, and
  Levental]{Diaz-Rohrer14}
Diaz-Rohrer, B.~B., K.~R. Levental, K.~Simons, and I.~Levental, 2014.
\newblock Membrane raft association is a determinant of plasma membrane
  localization.
\newblock \emph{Proceedings of the National Academy of Sciences}
  111:8500--8505.

\bibitem[Lin and London(2013)]{Lin13}
Lin, Q., and E.~London, 2013.
\newblock Altering Hydrophobic Sequence Lengths Shows That Hydrophobic Mismatch
  Controls Affinity for Ordered Lipid Domains (Rafts) in the Multitransmembrane
  Strand Protein Perfringolysin O.
\newblock \emph{J Biol Chem.} 288:1340--1352.

\bibitem[Niemela et~al.(2007)Niemela, Ollila, Hyvonen, Karttunen, and
  Vattulainen]{Niemela07}
Niemela, P.~S., S.~Ollila, M.~T. Hyvonen, M.~Karttunen, and I.~Vattulainen,
  2007.
\newblock Assessing the Nature of Lipid Raft Membranes.
\newblock \emph{PLoS Comput Biol} 3:1--9.

\bibitem[Cicuta et~al.(2007)Cicuta, Keller, , and Veatch]{Cicuta07}
Cicuta, P., S.~L. Keller, , and S.~L. Veatch, 2007.
\newblock Diffusion of Liquid Domains in Lipid Bilayer Membranes.
\newblock \emph{The Journal of Physical Chemistry B} 111:3328--3331.

\bibitem[Baumgart et~al.(2003)Baumgart, Hess, and Webb]{Baumgart03}
Baumgart, T., S.~T. Hess, and W.~W. Webb, 2003.
\newblock Imaging coexisting fluid domains in biomembrane models coupling
  curvature and line tension.
\newblock \emph{Nature} 425:821--824.

\bibitem[Franks and Lieb(1982)]{Franks82}
Franks, N.~P., and W.~R. Lieb, 1982.
\newblock Molecular mechanisms of general anaesthesia.
\newblock \emph{Nature} 300:487--93.

\bibitem[Oleksiuk et~al.(2011)Oleksiuk, Jakovljevic, Vladimirov, Carvalho,
  Paster, Ryu, Meir, Wingreen, Kollmann, and Sourjik]{Oleksiuk11}
Oleksiuk, O., V.~Jakovljevic, N.~Vladimirov, R.~Carvalho, E.~Paster, W.~S. Ryu,
  Y.~Meir, N.~S. Wingreen, M.~Kollmann, and V.~Sourjik, 2011.
\newblock Thermal Robustness of Signaling in Bacterial Chemotaxis.
\newblock \emph{Cell} 145:312--321.

\bibitem[Kidd et~al.(2015)Kidd, Young, and Siggia]{Kidd15}
Kidd, P.~B., M.~W. Young, and E.~D. Siggia, 2015.
\newblock Temperature compensation and temperature sensation in the circadian
  clock.
\newblock \emph{Proceedings of the National Academy of Sciences}
  112:E6284--E6292.

\bibitem[Daniels et~al.(2008)Daniels, Chen, Sethna, Gutenkunst, and
  Myers]{Daniels08}
Daniels, B.~C., Y.-J. Chen, J.~P. Sethna, R.~N. Gutenkunst, and C.~R. Myers,
  2008.
\newblock Sloppiness, robustness, and evolvability in systems biology.
\newblock \emph{Current Opinion in Biotechnology} 19:389 -- 395.

\bibitem[Smith et~al.(2008)Smith, Renden, and von Gersdorff]{Smith08}
Smith, S.~M., R.~Renden, and H.~von Gersdorff, 2008.
\newblock Synaptic vesicle endocytosis: fast and slow modes of membrane
  retrieval.
\newblock \emph{Trends in Neurosciences} 31:559 -- 568.


\bibitem[Thompson et~al.(1985)Thompson, Masukawa, and Prince]{Thompson85}
Thompson, S.~M., L.~M. Masukawa, and D.~A. Prince, 1985.
\newblock Temperature dependence of intrinsic membrane properties and synaptic
  potentials in hippocampal CA1 neurons in vitro.
\newblock \emph{J. Neurosci.} 5:817--24.

\bibitem[Tang et~al.(2010)Tang, Goeritz, Caplan, Taylor, Fisek, and
  Marder]{Tang10}
Tang, L.~S., M.~L. Goeritz, J.~S. Caplan, A.~L. Taylor, M.~Fisek, and
  E.~Marder, 2010.
\newblock Precise Temperature Compensation of Phase in a Rhythmic Motor
  Pattern.
\newblock \emph{PLoS Biol} 8:1--13.

\bibitem[Meyer(1901)]{Meyer01}
Meyer, H., 1901.
\newblock Zur Theorie der Alkoholnarkose. 3. Mittheilung: Der Einfluss
  wechselnder Temperatur auf Wirkungsstrke und Theilungscoefficient der
  Narcotiea.
\newblock \emph{Arch. exp. Path. Pharmak. (Naunyn-Schmiedebergs)} 46:338−346.

\bibitem[Cherkin and Catchpool(1964)]{Cherkin64}
Cherkin, A., and J.~F. Catchpool, 1964.
\newblock Temperature Dependence of Anesthesia in Goldfish.
\newblock \emph{Science} 144:1460--1462.

\bibitem[Barańska and Wlodawer(1969)]{Baranska69}
Barańska, J., and P.~Wlodawer, 1969.
\newblock Influence of temperature on the composition of fatty acids and on
  lipogenesis in frog tissues.
\newblock \emph{Comparative Biochemistry and Physiology} 28:553--570.


\bibitem[Anderson(1970)]{Anderson70}
Anderson, T.~R., 1970.
\newblock Temperature adaptation and the phospholipids of membranes in goldfish
  (carassius auratus).
\newblock \emph{Comparative Biochemistry and Physiology} 33:663--687.


\bibitem[Janoff et~al.(1981)Janoff, Pringle, and Miller]{Janoff81}
Janoff, A.~S., M.~J. Pringle, and K.~W. Miller, 1981.
\newblock Correlation of General Anesthetic Potency with Solubility in
  Membranes.
\newblock \emph{Biochimica Et Biophysica Acta} 649:125--128.

\bibitem[Bristow and Martin(1987)]{Bristow87}
Bristow, D.~R., and I.~L. Martin, 1987.
\newblock Solubilisation of the gamma-aminobutyric acid/benzodiazepine receptor
  from rat cerebellum: optimal preservation of the modulatory responses by
  natural brain lipids.
\newblock \emph{J Neurochem} 49:1386--93.

\bibitem[Sooksawate and Simmonds(2001)]{Sooksawate01b}
Sooksawate, T., and M.~Simmonds, 2001.
\newblock Effects of membrane cholesterol on the sensitivity of the GABAA
  receptor to GABA in acutely dissociated rat hippocampal neurones.
\newblock \emph{Neuropharmacology} 40:178 -- 184.


\end{thebibliography}


\end{document}

\newpage

\listoffigures

\begingroup
    \includegraphics[width=3.25in]{method_fig_v1.png}
		
\textcolor{black}{
\noindent {\bf Fig. 1.} Determination of the average critical temperature or pressure of DiIC$_{12}$ labeled GPMVs through fluorescence imaging. (A) Fields containing multiple GPMVs were imaged over a range of temperatures and at fixed pressure, with representative subsets of images shown on the left.  At high temperature, most GPMVs appear uniform, while an increasing fraction of vesicles appear phase separated as temperature is lowered, with phase separated vesicles indicated by yellow arrows.  We manually tabulate the fraction of GPMVs that contain two coexisting liquid phases as a function of temperature from these images, constructing the plot on the right.  These points are fit to the sigmoid function described in Methods to determine the extrapolated temperature where 50$\%$ of vesicles contain coexisting liquid phases.  (B) Fields containing multiple GPMVs were imaged over a range of pressures at fixed temperature, and representative subsets of images are shown on the left.  At low pressure, most GPMVs appear uniform, while an increasing fraction of vesicles appear phase separated as pressure is increased. As with the fixed pressure data in A, these points are fit to the sigmoid function described in Methods to determine the extrapolated pressure where 50$\%$ of vesicles contain coexisting liquid phases.  
}
\endgroup

\clearpage

\begingroup
    \includegraphics[width=3.75in]{figures/fig1v6.png}

\noindent {\bf Fig. 2.} Hexadecanol raises $T_c$ in GPMVs from rat (RBL) and \textit{Xenopus} (XtC-2) cell lines and can counteract the $T_c$ lowering effects of ethanol. (A) Values  indicate the average shift in $T_c$  ($\Delta T_c$) in a population of vesicles upon treatment with the compounds indicated. Solutions containing hexadecanol were prepared to be super-saturated as described in Methods.  Each point represents a single measurement and error bounds represent the 68\% confidence interval on the extrapolated $\Delta T_c$.   (B) Plots showing fraction of phase separated vesicles vs. temperature for the three points inside the gray box in A. 
\endgroup

\clearpage


\begingroup
    \includegraphics[width=6.75in]{figures/fig2v7.png}
    
\noindent {\bf Fig. 3.} (A) (Upper panel) Tadpole loss of righting reflex (LRR) for a titration of ethanol alone and combinations of ethanol and hexadecanol (EtOH+Hex) or ethanol and tetradecanol (EtOH+Tet) measured after 1h incubation in equilibrated solutions. At a given ethanol concentration,the presence of hexadecanol increases the fraction of tadpoles which respond to stimulus.  (Lower panel)  $\Delta T_c$ in RBL derived GPMVs for identical titrations of ethanol and EtOH+Hex. All solutions contain the ethanol concentration indicated by the lower horizontal axis. Red circle points additionally contain  hexadecanol concentrations indicated by the upper horizontal axis, and green triangle points additionally contain either 5 or 10$\mu$M tetradecanol.  (B) Time-course of LRR for one ethanol and ethanol+hexadecanol combination.  (C) (Left) Points in A replotted plotted as LRR vs $\Delta T_c$, including additional experiments with other n-alcohol combinations as indicated in the legend. (Center) LRR  plotted vs. aggregate hydrophobicity, tabulated by summing the concentration of each n-alcohol present normalized by its $AC_{50}$~\citep{Pringle81}, using $3\mu$M as a proxy $AC_{50}$ for hexadecanol and $5\mu$M as a proxy $AC_{50}$ for tetradecanol. (Right) LRR  plotted vs. the net anesthetic concentration, tabulated by summing the concentration of each known anesthetic present normalized by its $AC_{50}$~\citep{Pringle81}.  In each case the black line and purple shaded region denotes best fit and 50\% confidence intervals for model where LRR is quadratic in the respective x-axes.  The legend applies to all panels.
\endgroup


\clearpage

\begingroup
    \includegraphics[width=3.75in]{figures/fig3v1.png}

\noindent {\bf Fig. 4.} Ro15-4513 and DHM block the acute toxicity and intoxicating effect of ethanol. Each raise $T_c$ and cancel the effects of ethanol when added to GPMVs at the same concentration at which they are effective \textit{in vivo}. 
\endgroup

\clearpage

\begingroup
    \includegraphics[width=3.75in]{figures/fig4v1.png}

\noindent {\bf Fig. 5.} (A) The fraction of vesicles which are macroscopically phase separated is plotted as a function of hydrostatic pressure at three different temperatures, both for control vesicles and vesicles incubated with 12mM bBtOH.  In each case, increasing the pressure leads to an increase in the fraction of vesicles which are macroscopically phase separated.  (B)  $T_c$ is raised with increasing hydrostatic pressure in both control GPMVs and GPMVs incubated in butanol (BtOH). Here, 240$\pm$30 bar of hydrostatic pressure is required to reverse the effects of 12mM BtOH (shaded region). Closed symbols are obtained by extrapolating to find $T_c$ from data acquired at constant pressure while open symbols are obtained by extrapolating to find $P_c$ at constant temperature.  At temperatures above $T_c$ most vesicles are composed of a macroscopically uniform single liquid, while below $T_c$ most are separated into two co-existing liquid phases.
\endgroup

\clearpage

\begingroup
    \includegraphics[width=3.75in]{figures/fig6.png}

\textcolor{black}{
\noindent {\bf Fig. 6.} Anesthetics lower the transition temperature of a biologically tuned critical point which could lead to mis-regulation of ion channels and other membrane bound proteins.  (A) Schematic phase diagrams for the plasma membrane of an untreated and n-alcohol treated cell.  Guided by experiments~\citep{Veatch08} we hypothesize that the plasma membrane lies in the white region where lipids, proteins and other membrane components are well mixed macroscopically into a single two dimensional liquid phase.  However, due to their close proximity to the critical point (star), thermal fluctuations lead to relatively large domains enriched in particular components.  When cooled into the gray two phase region, GPMVs separate into two coexisting liquid phases termed liquid-ordered and liquid-disordered. Several n-alcohol general anesthetics lower the critical temperature of the membrane~\citep{Gray13}, changing the distance above the critical point, $T-T_c$. Here we also show that treatments which antagonize anesthetic action \textit{raise} critical temperatures reversing the effects of n-alcohols on $T_c$.  (B) Under normal conditions a hypothetical ion channel (large blue inclusion) has a tendency to inhabit relatively large domains enriched in particular lipids and proteins.  When the membrane is taken away from the critical point,  the structure of these domains is altered, possibly leading to changes in ion channel gating and function through a variety of mechanisms discussed in the text.
}
\endgroup




}
\end{document}